\newcommand*{\ATLASLATEXPATH}{latex/}
\documentclass[cernpreprint,texlive=2011,txfonts,UKenglish]{latex/atlasdoc}

\usepackage{\ATLASLATEXPATH atlaspackage}
\usepackage{\ATLASLATEXPATH atlasbiblatex}
\usepackage{breakcites}
\usepackage{multirow}

\usepackage{\ATLASLATEXPATH atlascontribute}

\usepackage{\ATLASLATEXPATH atlasphysics}

\usepackage{verbatim}

\graphicspath{{logos/}{figures/}}

\def\wp{\ensuremath{W^\prime}}

\def\zp{\ensuremath{Z^\prime}}

\def\mt{$m_{\mathrm T}$}

\def\pythia{{\sc Pythia}}

\def\geant{{\sc Geant}}

\def\vrap{{\sc VRAP}}

\def\sherpa{{\sc Sherpa}}

\def\photos{{\sc Photos}}

\def\powhegbox{{\sc Powheg-Box}}

\def\mcsanc{{\sc Mcsanc}}

\newcommand{\syspair}[2] { ${#1}$\%~(${#2}$\%) }

\def\wpssm{\ensuremath{W'_{\rm SSM}}} 
\def\xbr{\ensuremath{\sigma \times B}}
\def\wpe{\ensuremath{W'\rightarrow e \nu}}
\def\wpmu{\ensuremath{W'\rightarrow \mu \nu}}

\def\wpl{\ensuremath{W'\rightarrow \ell \nu}}

\def\mwp{\ensuremath{m_{W'}}}

\AtlasTitle{\boldmath Search for new resonances in events with one lepton and missing transverse momentum in $pp$ collisions at $\sqrt s = 13$~\tev\ with the ATLAS detector}

\author{The ATLAS Collaboration}

\date{\today}

\PreprintIdNumber{CERN-PH-2016-143}

\AtlasJournalRef{Phys. Lett. B 762 (2016) 334}
\AtlasDOI{10.1016/j.physletb.2016.09.040}

\AtlasAbstract{%
A search for $W^\prime$ bosons in events with one lepton (electron or muon) and missing transverse momentum is presented. 
The search uses 3.2 fb$^{-1}$ of $pp$ collision data collected at $\sqrt{s}~=~13$~\tev\ by the ATLAS experiment at the LHC in 2015. 
The transverse mass distribution is examined and no significant excess of events above the level expected from Standard Model processes is observed. 
Upper limits on the $W^\prime$ boson cross-section times branching 
ratio to leptons are set as a function of the $W^\prime$ mass.
Within the Sequential Standard Model $W^\prime$ masses below 4.07~\TeV\ are excluded at the 95\% confidence level. This extends the limit set using LHC data at $\sqrt{s}=8$~\TeV\ by around 800~\GeV. 
}

\hypersetup{pdftitle={ATLAS draft},pdfauthor={The ATLAS Collaboration}}

\begin{document}

\maketitle



\section{Introduction}
\label{sec:intro}
Many models of physics beyond the Standard Model (SM) predict the
existence of new spin-1 gauge bosons that could be discovered at the Large Hadron Collider (LHC). While the details of
the models vary, conceptually these particles are heavier versions of
the SM $W$ and $Z$ bosons and are generically called \wp\ and
\zp\ bosons. 

In this letter, a search for a \wp\ boson is presented using 3.2  \ifb\ of $pp$ collision data collected 
with the ATLAS detector in 2015 at a centre-of-mass energy of 13~\tev. 
The results 
are interpreted in the context of the
benchmark Sequential Standard Model (SSM), i.e. the extended gauge model described in Ref.~\cite{ssm}, in which the
couplings of the \wpssm\ to fermions are assumed to be identical to those
of the SM $W$ boson. The decay of the SSM \wp\ to SM bosons is not allowed and 
interference between the SSM \wp\ and the SM $W$ boson is neglected.
The search is conducted in the $\wp \to \ell \nu$ channel, where $\ell$ is an electron or a muon.
The signature is a charged lepton with 
high transverse momentum (\pT) and substantial missing transverse momentum (\met) due to the undetected neutrino. 
The discriminant to distinguish signal and background is the transverse mass
\begin{linenomath}
\begin{equation}
m_{\mathrm T} = \sqrt{2\pT\met(1-\cos \phi_{\ell\nu})},
\label{eq:mt}
\end{equation}
\end{linenomath}
where $\phi_{\ell\nu}$ is the angle between the lepton and \met{} in
the transverse plane.\footnote{ATLAS uses a right-handed coordinate system with its origin at the nominal interaction point (IP) in the centre of the detector and the $z$-axis along the beam pipe. The $x$-axis points from the IP to the centre of the LHC ring, and the $y$-axis points upward. Cylindrical coordinates $(r,\phi)$ are used in the transverse plane, $\phi$ being the azimuthal angle around the beam pipe. The pseudorapidity is defined in terms of the polar angle $\theta$ as $\eta=-\ln\tan(\theta/2)$.} The dominant background for the $\wp \rightarrow \ell \nu$ search is the high-\mt\ tail of the
charged-current Drell--Yan ($q\bar{q'}\rightarrow W \rightarrow \ell \nu$) process. 

Previous searches for \wpssm\ bosons in the $\wp \to e \nu$ and $\wp \to \mu \nu$ channels were carried out by both the ATLAS and CMS collaborations using the Run-1 data. The previous ATLAS analysis is based on data corresponding to an integrated luminosity of 20.3~\ifb\ taken at a centre-of-mass energy of $\sqrt{s} = 8$~\tev\ and sets a 95\% confidence level (CL) lower limit
on the \wpssm\ mass of 3.24~\TeV~\cite{atlas_8tev_pub}. The CMS Collaboration published a search using 19.7~\ifb\ of $\sqrt{s} = 8~\TeV$ data from 2012 which excludes
\wpssm\ masses below 3.28~\TeV\ at 95\%~CL~\cite{cms_8tev_pub}.

\section{ATLAS detector}
\label{sec:detector}
The ATLAS experiment \cite{atlas:detector} at the LHC is a multi-purpose particle detector with a forward-backward symmetric cylindrical geometry and a near $4 \pi$ 
coverage in solid angle. It consists of an inner tracking
detector (ID) surrounded by a thin superconducting solenoid providing a \SI{2}{\tesla} axial magnetic field, electromagnetic (EM) and hadronic calorimeters, and a muon spectrometer (MS). The inner tracking detector covers the pseudorapidity range $|\eta|<2.5$. It consists of a silicon pixel detector including the newly installed insertable B-layer \cite{ibl:detector, ibl:addendum}, followed by silicon microstrip, and transition radiation
tracking detectors. 
Lead/liquid-argon (LAr) sampling calorimeters provide EM energy measurements with high granularity. A hadronic (steel/scintillator-tile) calorimeter covers the central pseudorapidity range ($|\eta|<1.7$). The endcap and forward regions are instrumented with LAr calorimeters for both the EM
and hadronic energy measurements up to $|\eta|=4.9$. The muon spectrometer surrounds the calorimeters and is based on three large air-core toroid superconducting magnets with eight coils each. 
The field integral of the toroids ranges between 2.0 and 6.0 Tm for most of the detector.
It includes a system of precision tracking chambers, over $|\eta| < 2.7$, and fast detectors for triggering, over $|\eta| < 2.4$. A two-level trigger system is used to select events. The first-level trigger is implemented in hardware and uses a subset of the detector information. This is followed by a software-based trigger system that reduces the accepted event rate to about \SI{1}{\kilo\hertz}.

\section{Background and signal simulation}
\label{sec:MC}

Monte Carlo (MC) simulation samples are used to model
the expected signal and background processes, 
with the exception of data-driven background estimates for events in which one final-state jet or photon satisfies the electron or muon selection criteria.

The main background is due to the charged-current Drell--Yan (DY) process, 
generated at next-to-leading order (NLO) in QCD using \powhegbox\ v2~\cite{Alioli:2010xd} and the CT10 parton distribution functions (PDF)~\cite{CT10}, 
with \pythia~8.186~\cite{pythia8} to model parton showering and hadronisation. The same setup is used for the
neutral-current DY ($q\bar{q}\rightarrow Z/\gamma^*\rightarrow \ell \ell$) process. In both cases,
samples for all three lepton flavours are generated, and the 
final-state photon radiation (QED FSR) is handled 
by \photos~\cite{fsr_ref}.  
The DY samples are normalised as a function of mass to a next-to-next-to-leading order (NNLO) perturbative QCD (pQCD) 
calculation using \vrap~\cite{vrap} and the CT14NNLO PDF set~\cite{Dulat:2015mca}.
In addition, NLO electroweak (EW) corrections beyond QED FSR are calculated with \mcsanc~\cite{Bardin:2012jk,Bondarenko:2013nu} at LO in pQCD
as a function of mass.
In order to combine the QCD and EW terms,
the so-called additive approach is used where the EW corrections 
are added to the NNLO QCD cross-section prediction.


Backgrounds from \ttbar\ and single top-quark production are estimated at NLO using \powhegbox.  
These processes use the CT10 PDF set and are interfaced 
to \pythia~6.428~\cite{Sjostrand:2006za} for parton showering and hadronisation.   
Further backgrounds are due to diboson ($WW$, $WZ$ and $ZZ$) production.
These processes are generated 
with \sherpa~2.1.1 \cite{Gleisberg:2008ta} using the CT10 PDF set.

Signal samples for the $W^\prime \rightarrow e \nu$ and $W^\prime \rightarrow \mu \nu$ processes are produced at leading order (LO) in QCD
using \pythia~8.183 and the NNPDF2.3 LO PDF set.
The \wpssm\ boson has the same
couplings to fermions as the Standard Model $W$ boson and is assumed not to couple to the SM $W$ and $Z$ bosons. 
Interference effects between the \wp\ and the SM $W$ boson are neglected. In this model the branching ratio to a charged lepton and a neutrino is 8.2\% in the entire mass range considered in this search. The decay $W^\prime \rightarrow \tau \nu$, where the $\tau$~lepton subsequently decays leptonically is not treated as part of the signal. If included, this decay would constitute a very small contribution.  
The signal samples are normalised to the same mass-dependent NNLO pQCD calculation as used for the DY process.   
The EW corrections beyond QED FSR are not applied to the signal samples because
they depend on the couplings of the new particle to $W$ and $Z$ bosons, and are therefore model-dependent. 
The resulting cross-section times branching ratio for \wpssm\ masses of 2, 3 and 4~\tev\ are 153, 15.3 and 2.25~fb, respectively.

For all samples used in this analysis, the effects of multiple interactions per bunch crossing (“pile-up”) are
accounted for by overlaying simulated minimum-bias events. 
The interaction of particles with the detector and its response are modelled 
using a full \mbox{ATLAS} detector simulation~\cite{SOFT-2010-01} performed by \geant4~\cite{geant}.
Differences between data and simulation are accounted for in 
the lepton trigger, reconstruction, identification \cite{elepaper, muonpubnote}, 
and isolation efficiencies as well as the lepton energy/momentum resolution and 
scale \cite{elecalibpaper, muonpubnote}.

\section{Object reconstruction and event selection}
\label{sec:selection}
Events in the muon channel are selected by a trigger requiring that at least one muon with $\pt > 50$~\gev\ is found. These muons must be reconstructed in both the MS and the ID. 
In the electron channel, events are selected by a trigger requiring at least one electron candidate with $\pt > 24$~\gev\ that satisfies the medium identification criteria or a trigger requiring at least one electron with $\pt > 120$~\gev\ that satisfies the loose identification criteria. The selection cuts used to select electron candidates at trigger level are very similar to the ones used in the offline reconstruction and were optimised using a likelihood approach \cite{elepaper}.

The selected events must have a reconstructed primary vertex, which is the interaction vertex with the highest sum of $\pt^2$ of tracks found in the event. 
Each vertex reconstructed in the event consists of at least two associated tracks with $\pt > 0.4$~\gev. Only data taken during
periods when 
all detector components and the trigger readout are functioning well are considered.

Muons are reconstructed 
from MS tracks and matching ID tracks within $|\eta|<2.5$, requiring that the MS tracks have at least three hits
in each of the three separate layers of MS chambers to ensure optimal resolution for high-momentum muons \cite{muonpubnote}. 
In addition, these combined muons are required to pass a track quality selection based on the number of hits in the ID. To reduce sensitivity to the relative barrel--endcap alignment in the MS, the region $1.01~<~|\eta|~<~1.10$ is vetoed.
Muons are rejected if the difference between the muon charge-to-momentum ratios measured in the ID and MS exceeds seven times the sum in quadrature of the corresponding uncertainties, or if the track crosses poorly aligned MS chambers.
To ensure that the muons originate from the primary vertex, the transverse impact parameter significance, which is the ratio of the absolute value of the transverse impact parameter ($d_0$) to its uncertainty, has to be below three. 
The distance between the $z$-position of the point of closest approach of the muon track in the ID to the beamline and the $z$-coordinate of the primary vertex is required to be less than $10$~mm. 
Furthermore, only isolated muons are considered. The scalar sum over the track \pt\ in an isolation cone around the muon (excluding the muon itself) divided by the muon \pt\ is required to be below a \pt-dependent cut tuned for a 99\% efficiency. The isolation cone size $\Delta R = \sqrt{(\Delta \eta)^2 + (\Delta \phi)^2}$ is defined as $10 \gev$ divided by the muon \pt\ and has a maximum size of $\Delta R = 0.3$. 

Electrons are
formed from clusters of cells in the electromagnetic calorimeter associated with a track in the ID. The electron \pt\ is obtained from the calorimeter energy
measurement and the direction of the associated track. 
The electron 
must be within the range $|\eta| < 2.47$ and outside the transition region between the barrel and endcap calorimeters ($1.37 < |\eta| < 1.52$). In addition, tight identification criteria \cite{elepaper} need to be satisfied. The identification uses a likelihood discriminant based on measurements of calorimeter shower shapes and measurements of track
properties from the ID.
 To ensure that the electrons originate from the primary vertex, the transverse impact parameter significance must be below five. 
Furthermore, calorimeter- and track-based isolation criteria, tuned for an overall efficiency of 98\%, independent of \pt, are applied. The sum of the calorimeter transverse energy deposits in the isolation cone of size $\Delta R = 0.2$ (excluding the electron itself) divided by the electron \pt is used in the discrimination criterion. The track-based isolation is determined similarly to that for muons. The scalar sum of the \pt\ of all tracks in a cone around the electron, divided by the electron \pt has to be below a given value.
The cone has a size $\Delta R = 10 \gev/p_{\mathrm T}(e)$ with a maximum value of $\Delta R = 0.2$.
 
The calculation of the missing transverse momentum is based on the selected electrons, photons, tau leptons, muons and jets found in the event. The value of \met{} is evaluated by the vector sum of the \pt\ of the physics objects selected in the analysis and the tracks not belonging to any of these physics objects \cite{etmisspaper}.
Jets used in the \met{} calculation are reconstructed from clusters of calorimeter cells with $|\eta| < 5$ using the anti-$k_t$ algorithm \cite{antikt} with a radius parameter of $0.4$.
They are calibrated using the method described in Ref.~\cite{topo2} and are required to have $\pt~>~20$~\gev.

Events are selected if they have exactly one electron or muon with $\pt > 55$~\gev. The \met{} value found in the event is required to exceed 55~\gev\
and the transverse mass has to satisfy $m_{\mathrm T} > 110$~\gev. 
For these selection cuts the acceptance times efficiency, defined as the fraction of simulated candidate events that pass the event selection, amounts to 81\% (75\%) for the electron channel and 53\% (50\%) for the muon channel at a \wp\ mass of 2~\tev\ (4~\tev).

\section{Background estimate and comparison to data}
\label{sec:bg}

The background from processes with at least one prompt final-state lepton is estimated with simulated events. The processes with non-negligible contributions are charged-current DY ($W$ production), \ttbar and single top-quark production, in the following referred to as
``top-quark'' background, as well as neutral-current DY ($Z/\gamma^*$ production) and 
diboson production.

Background contributions from events where one final-state jet or photon passes 
the lepton
selection criteria are determined using a data-driven ``matrix'' method.
This includes contributions from multijet, heavy-flavour quark and $\gamma$ + jet production, 
referred to hereafter as the multijet background. 
The first step of the matrix method is to calculate 
the factor $f$, 
the fraction
of lepton candidates that pass the nominal lepton identification and isolation requirements
in a background-enriched data sample containing ``loose'' lepton candidates.
These loose candidates satisfy only a subset of the nominal criteria, which are
stricter than the trigger requirements imposed. 
Potential contamination of prompt final-state leptons in the background-enriched sample
is accounted for using MC simulation.
In addition to the factor $f$,
the fraction of
real leptons $r$ in the sample
of loose objects satisfying the nominal requirements is used in evaluating this background. 
This probability is computed from MC simulation.

The contribution to the background from events with a fake lepton is determined in the following way.
The relation between the number of 
real prompt leptons ($N_R$) or fake leptons ($N_F$)
and the number of measured objects 
found in the events containing the loose lepton candidates
($N_T, N_L$) can be written as
\begin{linenomath} 
\begin{equation}
\label{ff_matrix_full}
\begin{pmatrix} N_T \\ N_L \end{pmatrix}
=
  \begin{pmatrix}
  r &  f  \\
   (1-r)  & (1-f)
  \end{pmatrix}
  \begin{pmatrix} N_R \\ N_F \end{pmatrix},
\end{equation} 
\end{linenomath} 
where the subscript $T$ refers to leptons that pass the nominal selection. The subscript $L$ corresponds to
leptons that pass the loose requirements described above but
fail the nominal requirements. 
The number of jets and photons misidentified as leptons ($N_{T}^{\rm{Multijet}}$) in the total number of objects passing the signal selection 
($N_T$) is given as
\begin{linenomath} 
\begin{equation}
\label{bkg_true_quantities}
N_{T}^{\rm{Multijet} }  = f \, N_{F} = \frac{f}{r-f} \Big(\, r\,  (N_L + N_T) - N_T \Big).
\end{equation}
\end{linenomath} 
The right-hand side of Eq.~(\ref{bkg_true_quantities})
is obtained by solving Eq.~(\ref{ff_matrix_full}).

The simulated top-quark and diboson samples as well as the data-driven background estimate are statistically 
limited at large \mt. Therefore, the expected number of events is extrapolated into the high-\mt\
region using paramterisations of the \mt\ shape fitted to the expected background in the low-\mt\ region.
Several fits
are carried out based on the functions $f (m_{\mathrm T}) = a \, m_{\mathrm T}^b \, m_{\mathrm T}^{c \log m_{\mathrm T}}$ and $f (m_{\mathrm T}) = a/ (m_{\mathrm T} + b)^c$. These fits explore various fit ranges typically starting between 140 and 200~\GeV\ and extending up to 600 to 900~\GeV.
The fit with the best $\chi^2$ per degree of freedom 
is used as the extrapolated background contribution, with an uncertainty evaluated using the envelope of all performed fits.

Finally, the expected number of background events is calculated as the sum of the data-driven and 
simulated background estimates. The background is dominated by the charged-current DY production 
for all values of \mt, as can be seen in the upper panel of Figure~\ref{fig:mT}. 
For example, the contribution from charged-current DY is about 90\% for both channels at \mt\ $>1$~\TeV.
In both channels, the number of observed events agrees with the background estimate, as shown in the upper two panels of 
Figure~\ref{fig:mT} and in Table~\ref{tab:backgroundTable}.  
As can be seen in the middle panels, 
the data are systematically above the predicted background at low \mt\ but
are within the $\pm 1\sigma$ uncertainty band, which is dominated by 
the \met\ related systematic uncertainties in this region. 
The lower panels of Figure~\ref{fig:mT} show the ratio of the data to 
the adjusted background that results from the statistical analysis described in Section~\ref{sec:result}. The data agree well with 
the adjusted background prediction.

\begin{figure}
  \centering
  \includegraphics[width=0.495\columnwidth]{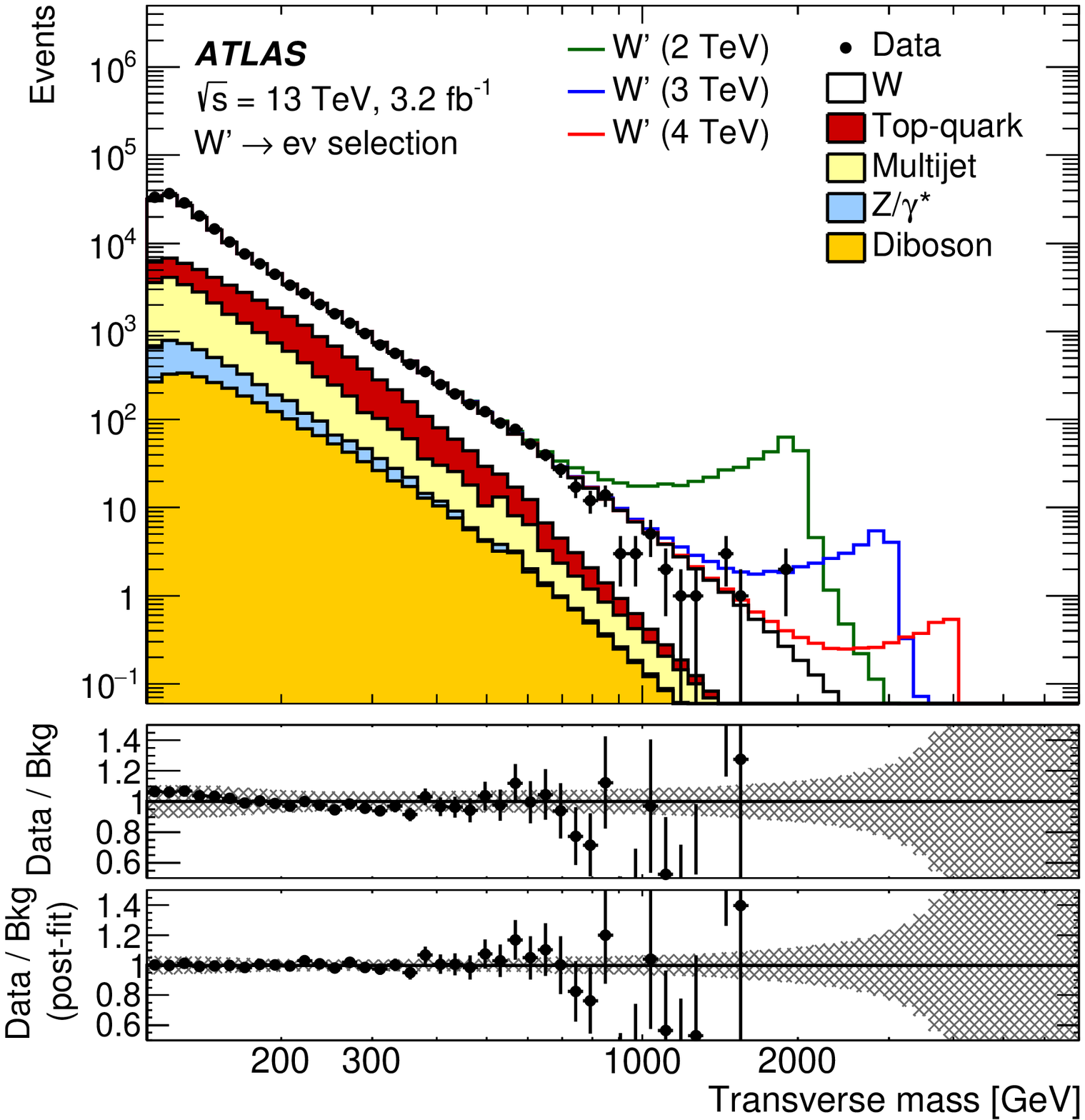}
  \includegraphics[width=0.495\columnwidth]{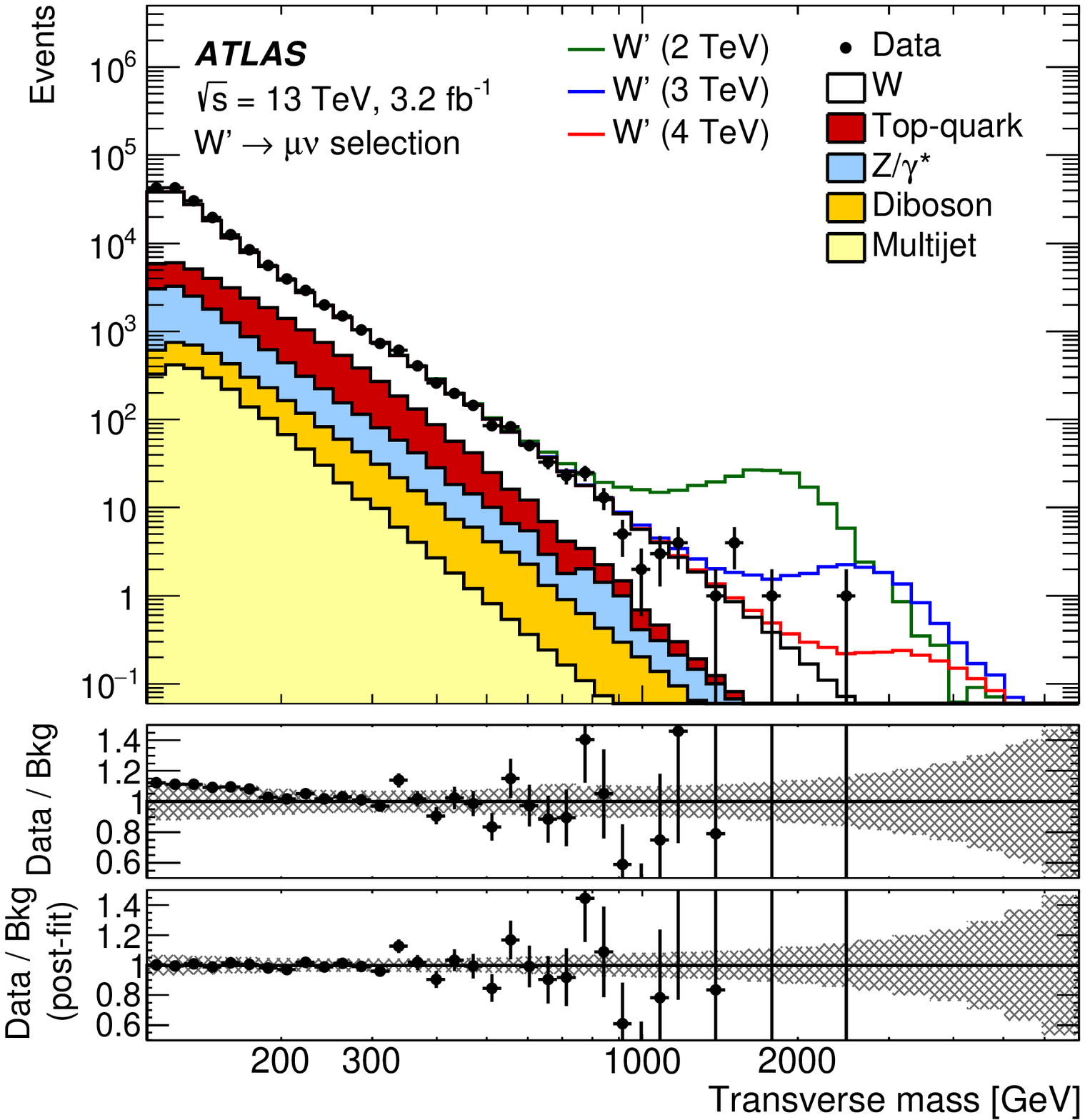}
  \caption{
Transverse mass distributions for events satisfying all selection criteria in the electron (left) and muon (right) channels.
The distributions are compared to the stacked sum of all expected backgrounds,
with three selected \wpssm\ signals overlaid. 
The bin width is constant in $\log m_{\mathrm T}$. 
The middle panels show the ratio of the data to the expected background.
The lower panels show the ratio of the data to the adjusted expected background (``post-fit'') that results from the statistical analysis.
The bands in the ratio plots indicate the sum in quadrature of the systematic uncertainties.
}
  \label{fig:mT}
\end{figure}
\begin{table*}[htbp]
\fontsize{3mm}{3.5mm}\selectfont
\setlength{\abovecaptionskip}{6pt}
\setlength{\belowcaptionskip}{6pt}
\caption{The expected and observed numbers of events in the electron (top) and muon (bottom) channels
in bins of \mt. The errors quoted are the combined statistical and systematic uncertainties. 
The systematic uncertainty includes all systematic uncertainties 
except the one for the integrated luminosity (5\%).
}
\label{tab:backgroundTable}
\begin{tabular}{l r@{ $\pm$ }l r@{ $\pm$ }l r@{ $\pm$ }l r@{ $\pm$ }l r@{ $\pm$ }l r@{ $\pm$ }l r@{ $\pm$ }l }
\multicolumn{15}{c}{Electron Channel} \\
\hline
\hline
$m_\mathrm{T}$ [\gev] & \multicolumn{2}{c}{110--150} & \multicolumn{2}{c}{150--200} & \multicolumn{2}{c}{200--400} & \multicolumn{2}{c}{400--600} & \multicolumn{2}{c}{600--1000} & \multicolumn{2}{c}{1000--3000} & \multicolumn{2}{c}{3000--7000} \\
\hline
Total SM	&122000 & 11000	&32600 & 2100	&14700 & 600	&845 & 34	&167 & 9	&19.1 & 1.5	&0.0261 & 0.0032\\
\hline
SM + \wp\ (2 \tev)	&122000 & 11000	&32600 & 2100	&14700 & 600	&864 & 35	&223 & 9	&344.8 & 2.7	&0.100 & 0.005\\
SM + \wp\ (3 \tev)	&122000 & 11000	&32600 & 2100	&14700 & 600	&847 & 34	&170 & 9	&50.7 & 1.7	&2.150 & 0.100\\
SM + \wp\ (4 \tev)	&122000 & 11000	&32600 & 2100	&14700 & 600	&846 & 34	&167 & 9	&21.4 & 1.5	&2.013 & 0.018\\
SM + \wp\ (5 \tev)	&122000 & 11000	&32600 & 2100	&14700 & 600	&846 & 34	&167 & 9	&19.5 & 1.5	&0.331 & 0.004\\
\hline
\hline
Data	&\multicolumn{2}{c}{129497}	&\multicolumn{2}{c}{32825}	&\multicolumn{2}{c}{14260}	&\multicolumn{2}{c}{846}	&\multicolumn{2}{c}{149}	&\multicolumn{2}{c}{15}	&\multicolumn{2}{c}{0}\\
\hline
\hline
& \multicolumn{2}{c}{} & \multicolumn{2}{c}{} & \multicolumn{2}{c}{} & \multicolumn{2}{c}{} & \multicolumn{2}{c}{} & \multicolumn{2}{c}{} & \multicolumn{2}{c}{} \\ 
\multicolumn{15}{c}{Muon Channel} \\
\hline
\hline
$m_\mathrm{T}$ [\gev] & \multicolumn{2}{c}{110--150} & \multicolumn{2}{c}{150--200} & \multicolumn{2}{c}{200--400} & \multicolumn{2}{c}{400--600} & \multicolumn{2}{c}{600--1000} & \multicolumn{2}{c}{1000--3000} & \multicolumn{2}{c}{3000--7000} \\
\hline
Total SM &   118000 & 12000  &   29700 & 2600  &   12100 & 600  &   660 & 40  &   135 & 11  &   14.6 & 1.4  &   0.058 & 0.013  \\ 
\hline 
SM + \wp\ (2 \tev) &   118000 & 12000  &   29700 & 2600  &   12100 & 600  &   670 & 40  &   175 & 13  &   214 & 16  &   2.0 & 0.8  \\ 
SM + \wp\ (3 \tev) &   118000 & 12000  &   29700 & 2600  &   12100 & 600  &   660 & 40  &   137 & 11  &   31.8 & 2.5  &   3.8 & 0.4  \\ 
SM + \wp\ (4 \tev) &   118000 & 12000  &   29700 & 2600  &   12100 & 600  &   660 & 40  &   135 & 11  &   16.2 & 1.5  &   1.16 & 0.11  \\ 
SM + \wp\ (5 \tev) &   118000 & 12000  &   29700 & 2600  &   12100 & 600  &   660 & 40  &   135 & 11  &   14.9 & 1.4  &   0.227 & 0.025  \\
\hline
\hline
Data & \multicolumn{2}{c}{131672} & \multicolumn{2}{c}{31980} & \multicolumn{2}{c}{12393} & \multicolumn{2}{c}{631} & \multicolumn{2}{c}{121} & \multicolumn{2}{c}{15} & \multicolumn{2}{c}{0} \\
\hline
\hline
\end{tabular}
\end{table*}

\section{Systematic uncertainties}
\label{sec:systematics}
\begin{table}
\caption{Systematic uncertainties in the expected number of events as evaluated at \mt\ $=$ 2 (4)~\TeV, both for signal events 
with a \wpssm\ mass of 2~(4)~\TeV\ and for background. 
Uncertainties that are not applicable are denoted ``{\sc n/a}''. \label{tab:syst}}
\begin{center}
\centering
\small
\begin{tabular}{l|cc|cc}
\toprule
\toprule
Source  &  \multicolumn{2}{c|}{Electron channel}  &  \multicolumn{2}{c}{Muon channel}  \\
&  Background  &  Signal  &  Background  &  Signal  \\
\midrule
Trigger &\syspair{1}{<0.5} & \syspair{1}{<0.5} &\syspair{3}{4} & \syspair{4}{4}\\
Lepton reconstruction &\multirow{2}{*}{\syspair{3}{3}} & \multirow{2}{*}{\syspair{3}{3}} &\multirow{2}{*}{\syspair{5}{8}} & \multirow{2}{*}{\syspair{5}{7}}\\
and identification & & & & \\
Lepton isolation &\syspair{2}{2} & \syspair{2}{2} &\syspair{5}{5} & \syspair{5}{5}\\
Lepton momentum &\multirow{2}{*}{\syspair{4}{6}} & \multirow{2}{*}{\syspair{10}{7}} &\multirow{2}{*}{\syspair{3}{11}} & \multirow{2}{*}{\syspair{1}{4}}\\
scale and resolution & & & & \\
\met{} resolution and scale &\syspair{<0.5}{<0.5} &\syspair{<0.5}{<0.5} &\syspair{<0.5}{<0.5} &\syspair{<0.5}{<0.5}\\
Jet energy resolution &\syspair{<0.5}{<0.5} &\syspair{<0.5}{<0.5} &\syspair{1}{2} &\syspair{<0.5}{<0.5}\\
\midrule
Multijet background &\syspair{2}{15} & {\sc n/a} ({\sc n/a}) & \syspair{1}{1} & {\sc n/a} ({\sc n/a})\\
Diboson \& top-quark bkg. &\syspair{6}{49} & {\sc n/a} ({\sc n/a}) &\syspair{5}{15} & {\sc n/a} ({\sc n/a})\\
PDF choice for DY &\syspair{1}{22} & {\sc n/a} ({\sc n/a}) &\syspair{<0.5}{1} & {\sc n/a} ({\sc n/a})\\
PDF variation for DY &\syspair{9}{19} & {\sc n/a} ({\sc n/a}) &\syspair{8}{12} & {\sc n/a} ({\sc n/a})\\
Electroweak corrections &\syspair{5}{9} & {\sc n/a} ({\sc n/a}) &\syspair{4}{6} & {\sc n/a} ({\sc n/a})\\
\midrule
Luminosity &\syspair{5}{5} &\syspair{5}{5} &\syspair{5}{5} &\syspair{5}{5}\\
\midrule
Total &\syspair{14}{60} & \syspair{11}{8} &\syspair{14}{25} & \syspair{9}{12}\\
\bottomrule
\bottomrule
\end{tabular}
\end{center}

\end{table}

Experimental systematic uncertainties arise from the background and luminosity estimates, the trigger selection, the lepton reconstruction, 
identification and isolation criteria \cite{elepaper, muonpubnote}, as well as effects of the 
energy/momentum scale and resolution \cite{elecalibpaper, muonpubnote}. 
The systematic uncertainties for the two channels are summarised in Table~\ref{tab:syst}.
At large \mt, the dominant source of uncertainty is due to the background extrapolations 
in the electron and muon channels, described in Section~\ref{sec:bg}, and to the momentum resolution in the muon channel.  
The extrapolation uncertainties are shown in Table~\ref{tab:syst} for the data-driven 
multijet background and the combined top-quark and diboson backgrounds.
The multijet background uncertainty in the electron channel includes a 25\% contribution from the data-driven estimate, 
which is due to the dependence of the factor $f$ (see Section~\ref{sec:bg}) on the specific selection used to derive the
background-enriched sample. No additional uncertainty is assigned in the muon case as the multijet background 
is small. 

The electron and muon reconstruction, identification and isolation efficiencies as well as their corresponding uncertainties
were evaluated from data using tag-and-probe methods in $Z$ boson decays up to a $\pt$ of $\mathcal{O}$(100 \gev).  
The ratio of the efficiency measured in data to that of the MC simulation is then used to correct 
the MC prediction. 
For electrons, these ratios are measured following the prescriptions of Ref.~\cite{eleperf}, with adjustments for the
2015 running conditions.
For higher-\pt\ electrons, an additional systematic uncertainty of 2.5\% is assigned to the identification efficiency. This is 
based on differences observed between data and simulation, and their propagation to the simulated electrons.
For the isolation efficiency, an additional uncertainty of 2\% is attributed to high-\pt\ electrons 
from the variation of the mean values of the ratio of the isolation efficiencies between data and simulation in various \pt\ and \eta\ bins.
For muons, no significant dependence of the ratio of the efficiencies measured in data over the ones measured in MC simulation 
as a function of \pt\ is observed~\cite{muonpubnote}. 
For high-\pt\ muons an upper limit on the uncertainty of 2–3\% per \tev\ is extracted 
from simulation. 
For the isolation criterion
an extrapolation of the uncertainties from the low-\pt\ muons is used and results in a 5\% uncertainty.

The systematic uncertainties related to \met{}\ originate from both the calculation of the contribution of tracks not associated with any 
physics object in the \met\ calculation \cite{etmisspaper} and the jet energy 
scale and resolution uncertainties \cite{topo2}. 
The uncertainties due to the jet energy and \met{}\ resolutions are small at large \mt, but have non-negligible contributions at small \mt, while 
the jet energy scale uncertainties are found to be negligible.

The uncertainty in the integrated luminosity is 5\%, affecting all simulated samples. It is derived following a methodology similar to that detailed in Ref.~\cite{lumipaper}, from a preliminary calibration of the luminosity scale, using a pair of $x$--$y$ beam-separation scans performed in August 2015.

Uncertainties on the theoretical aspects of the calculations for the background processes are considered, while for the \wp\ boson signal only the experimental uncertainties described above are evaluated.
They are related to the production cross-sections of the various backgrounds estimated from MC simulation.
The dominant uncertainty arises from the PDF for the charged-current DY background,
where the impact is larger in the electron channel 
than in the muon channel.
This is due to the better energy resolution in the electron channel, which leads to smaller migration of events from low \mt, 
where the PDF is better constrained, to high \mt. 
The PDF uncertainty
is obtained from the 90\%~CL CT14NNLO PDF error set, 
using \vrap\ in order to calculate the NNLO Drell--Yan cross-section as a function of mass. 
Instead of calculating only one overall PDF uncertainty based on the full set of 56 eigenvectors,
this analysis uses a reduced set of seven eigenvectors with a mass dependence similar to the one provided by the authors of the CT14 PDF using 
MP4LHC~\cite{gao:2013bia, butterworth:2015oua}. 
Their sum in quadrature is shown as ``PDF variation'' in Table~\ref{tab:syst}.
An additional uncertainty is assigned to account for
potential differences when using the 
MMHT2014~\cite{Harland-Lang:2014zoa} or NNPDF3.0~\cite{Ball:2014uwa} PDF sets.
Of these, only the central values for NNPDF3.0 fall outside the ``PDF variation'' uncertainty at large \mt.
Thus, an envelope of the 
``PDF variation'' and the NNPDF3.0 central value is formed, 
where 
the former is subtracted in quadrature from this envelope, 
and the remaining part, which is only non-zero when the NNPDF3.0 central value is outside the ``PDF variation'' uncertainty,
is quoted as ``PDF choice''.

Uncertainties in the higher order electroweak corrections are determined as the difference between the additive approach and a
factorised approach, which approximately span the range allowed for mixed EW and QCD contributions.
Uncertainties due to higher-order QCD corrections on the charged-current DY are estimated using VRAP by varying the renormalisation and factorisation scales 
simultaneously up and down by a factor of two and are found to be negligible.
Similarly, the uncertainty due to the imperfect knowledge of \alphas, obtained by varying \alphas\ by as much as 0.003 at large masses, can be neglected.

The $t\bar{t}$ MC sample is normalised to a cross-section 
of 
$\sigma_{t\bar{t}}= 832^{+20}_{-29}$ (scale) $ \pm 35$ (PDF + \alphas) pb as 
calculated 
with the Top++2.0 program and is accurate to NNLO in pQCD, including soft-gluon resummation 
to next-to-next-to-leading-log order (see Ref.~\cite{Czakon:2011xx} and references therein), and assuming a top-quark mass of 172.5 \gev. 
The first uncertainty comes from the independent variation of the factorisation and renormalisation scales, 
$\mu_{\mathrm F}$ and $\mu_{\mathrm R}$, 
while the second one is associated with variations in the PDF and \alphas, following 
the PDF4LHC prescription (see Ref.~\cite{Botje:2011sn} and references therein) with the MSTW2008 68\% CL NNLO~\cite{Martin:2009bu}, 
CT10 NNLO~\cite{Gao:2013xoa} and NNPDF2.3 NNLO~\cite{Ball:2012cx} PDF sets.
Normalisation uncertainties in the top-quark background are found to add a negligible contribution to the total background uncertainty.
The modelling of the top-quark background is verified in a data control region defined by requiring the presence of an additional muon (electron) in events passing the electron (muon) selection. The uncertainty in the diboson background is found to contribute negligibly to the total background uncertainty.


\section{Results}
\label{sec:result}

To test for excesses in data, a log-likelihood ratio test is carried out using RooStats \cite{roostat} to calculate 
the probability that the background 
fluctuates such as to give a signal-like
excess equal to or larger than what is observed. 
The likelihood functions are defined as the product of Poisson probabilities over all \mt\ bins in the search 
region (110 \gev < \mt\ < 7000 \gev) and Gaussian constraints for the nuisance parameters.
They are maximised for two cases: the presence of a signal 
above background, and background only.
The signal is modelled using \wpssm\ templates binned in \mt\ for a series
of \wpssm\ masses covering the full considered mass range. As examples, three of these templates are shown in Figure~\ref{fig:mT}
for both channels. 

As no excess more significant than 2$\sigma$ is observed in the log-likelihood ratio test,
upper limits on the cross-section for the production of a new boson times its
branching ratio to only one lepton generation (\xbr) are determined at 95\% CL as a function of the mass of the boson, $m_{W^\prime}$.
The observed upper limits are derived by comparing 
data to the expected background,
using templates for the shape of the simulated \mt\ distributions for different signal masses. 
Similarly, 
the expected limit is determined using pseudo-experiments obtained from the estimated background distributions, instead of the actual data.
The pseudo-experiments result in a distribution of limits, 
the median of which is taken as the expected limit, and $\pm 1\sigma$ and $\pm 2\sigma$ bands are defined as the ranges containing 
respectively 68\%\ and 95\%\ of the limits obtained with the pseudo-experiments.
The limit setting is based on a Bayesian approach detailed in Ref.~\cite{EXOT-2012-23}, using the Bayesian Analysis Toolkit~\cite{bayes}, 
with a uniform positive prior probability distribution for \xbr.

Figure~\ref{fig:e_mu_limits} 
presents the expected and observed limits separately for the electron and muon 
channels. 
Figure~\ref{fig:comb_limits} shows their combination, taking into account that the
theoretical uncertainties as well as the systematic uncertainties in the \met{}, jet energy resolution and luminosity are correlated between the channels.
\begin{figure}[htbp]
  \centering
  \includegraphics[width=0.495\columnwidth]{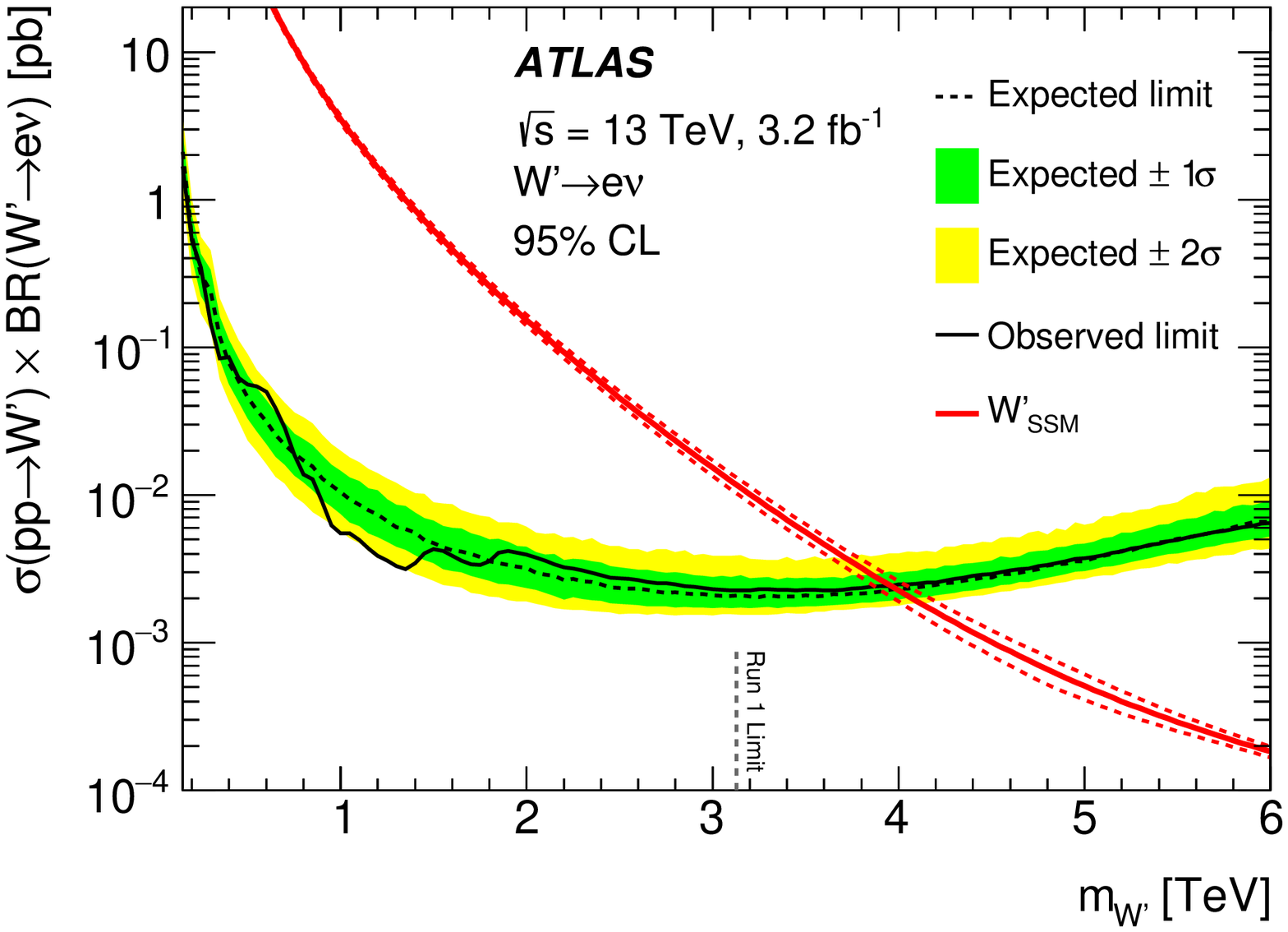}
  \includegraphics[width=0.495\columnwidth]{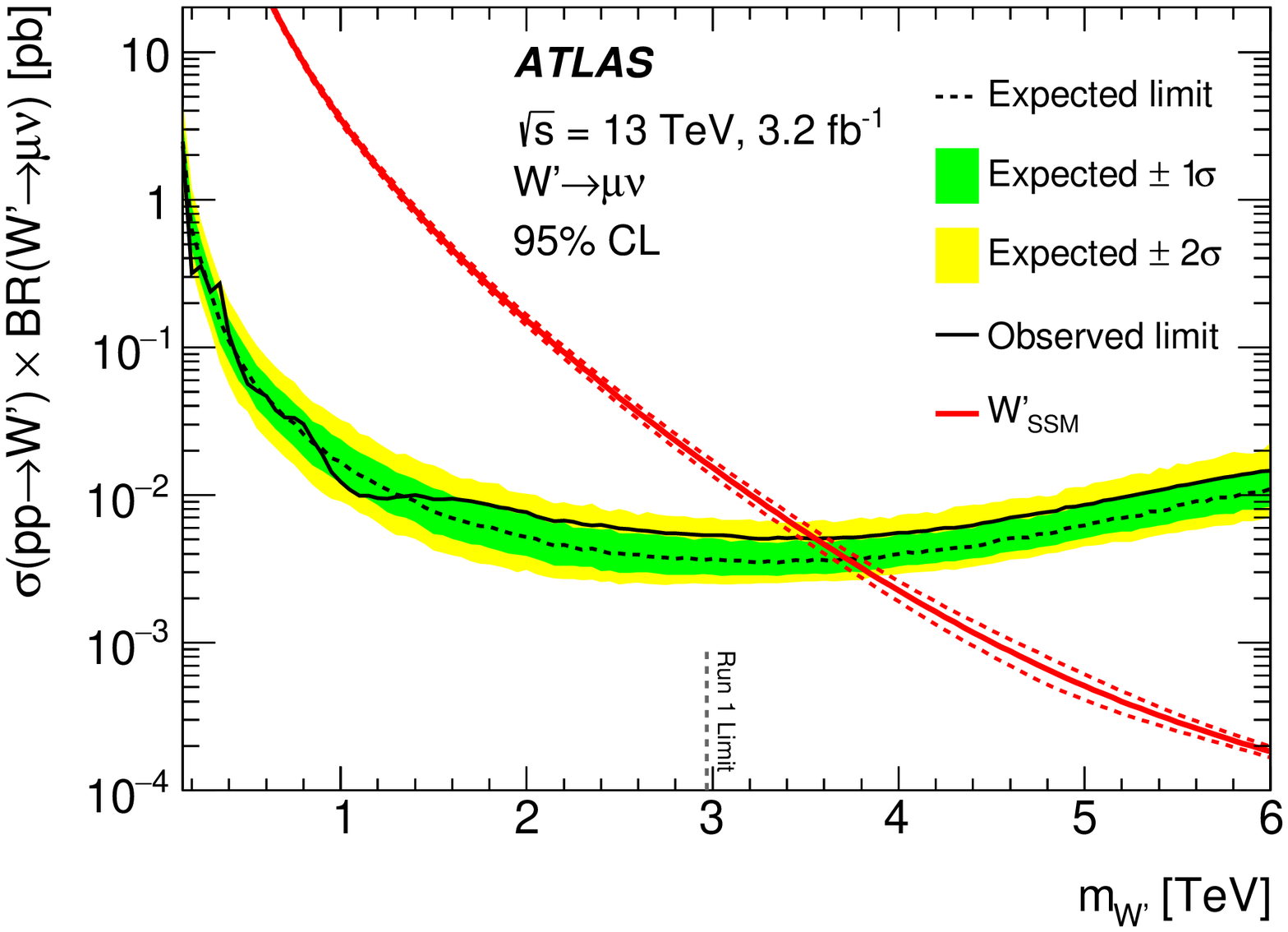}
  \caption{
Median expected (dashed black line) and observed (solid black line) 95\%~CL upper limits on cross-section times branching ratio (\xbr) in 
the electron (left) and muon (right) channels.
The bands for 68\% (green) and 95\% (yellow) confidence intervals are also shown.
The predicted \xbr\ for \wpssm\ production is shown as a red solid line. Uncertainties in \xbr\ from the PDF, \alphas\ and scale are shown as a red-dashed line.
The vertical dashed line indicates the mass limit of the 8~\tev\ data analysis \cite{atlas_8tev_pub}.
}
  \label{fig:e_mu_limits}
\end{figure}
\begin{figure}[htbp]
  \centering
  \includegraphics[width=0.65\columnwidth]{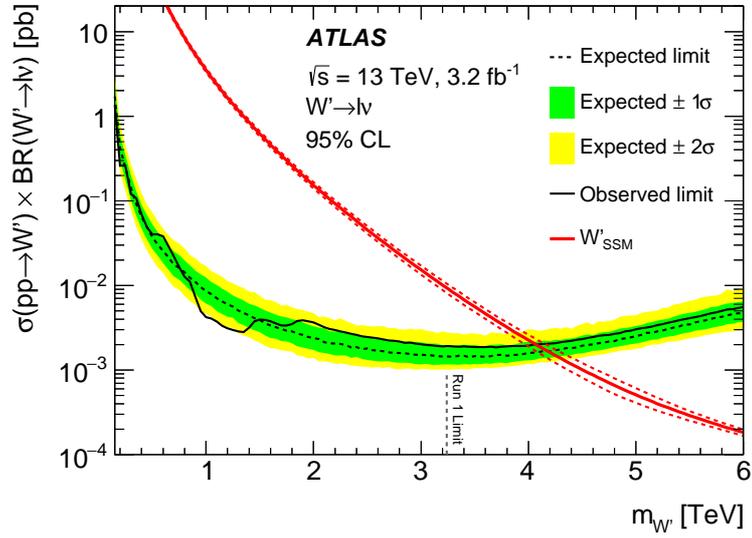}
  \caption{
Median expected (dashed black line) and observed (solid black line) 95\%~CL upper limits on cross-section times 
branching ratio (\xbr) in the combined channel, along with predicted \xbr\ for \wpssm\ production (red line). 
Uncertainties in \xbr\ from the PDF, \alphas\ and scale are shown as a red-dashed line. 
The bands for 68\% (green) and 95\% (yellow) confidence intervals are also shown.
The vertical dashed line indicates the mass limit of the 8~\tev\ data analysis \cite{atlas_8tev_pub}.
}
  \label{fig:comb_limits}
\end{figure}
The expected upper limit on \xbr\ is stronger in the electron channel due to the larger acceptance times efficiency 
and the better momentum resolution (see Section \ref{sec:selection}). 
The difference in resolution can be seen in Figure~\ref{fig:mT} when comparing the shapes of the three 
reconstructed \wpssm\ signals. 
For both channels and their combination, the observed limit 
does not deviate above the $2\sigma$ band of expected limits for all $m_{W^\prime}$.

For specific models with a known \xbr\ as a function of mass, the upper limit on \xbr\ 
can be used to set a lower mass limit on the new resonance, e.g. for the benchmark \wpssm\ model. 
Figures \ref{fig:e_mu_limits} and \ref{fig:comb_limits} show the predicted \xbr\ for the \wpssm\ as a function of its mass.
Uncertainties on \xbr\ from the PDF, \alphas\ and scale are shown as a red-dashed line.
The resulting expected and observed lower limits on the \wpssm\ mass are given in Table~\ref{tab:limits_mass_wp}.
The observed limit in the muon and in the combined channel is weaker than the expected one due to a few events in the muon 
channel above approximately 1.5~\tev\ in \mt, as can be seen in the right panel of Figure~\ref{fig:mT}.
\begin{table}[!htbp]
  \centering
  \begin{tabular}{c|cc}
    \hline
    \hline
    &  \multicolumn{2}{c}{\mwp\ lower limit [\TeV]} \\
    Decay     &  Expected & Observed \\
    \hline
    \wpe  & 3.99 & 3.96 \\
    \wpmu & 3.72 & 3.56 \\
    \wpl  & 4.18 & 4.07 \\
    \hline
    \hline
  \end{tabular}
  \caption{Expected and observed 95\% CL lower limit on the \wpssm\ mass in the electron and muon channels and their combination.}
  \label{tab:limits_mass_wp}
\end{table}

To compare to previous ATLAS searches, the cross-section limits for \wp\ bosons normalised to the SSM predictions as a function of mass are displayed in Figure~\ref{fig:combinedlimits}. The limit based on the 13~\tev\ data is similar to the 8~\tev\ data limit in the mass range between 0.5 and 2.3~\tev. At lower and higher mass values, the new limit improves compared to the previous results.
\begin{figure}[hbtp]
  \centering
  \includegraphics[width=0.7\textwidth]{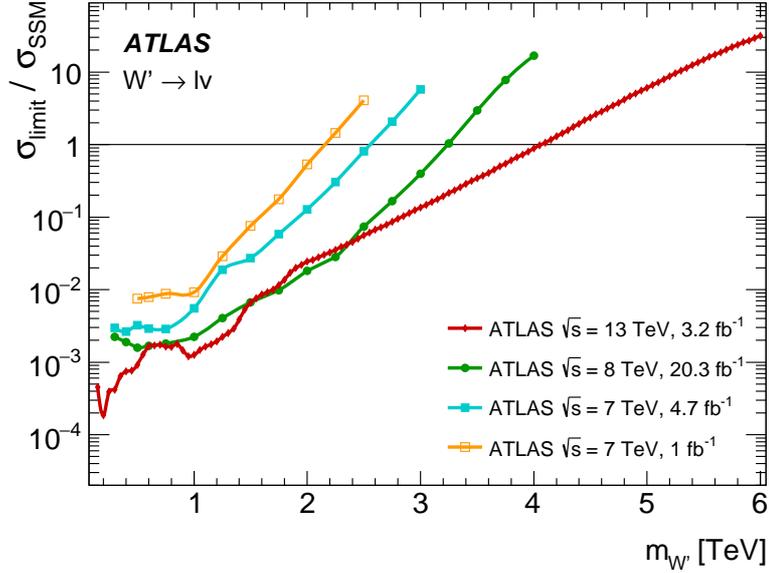}
  \caption{Normalised cross-section limits ($\sigma_{\mathrm{limit}}/\sigma_{\mathrm{SSM}}$) for \wp\ bosons as a function of mass for this analysis and from previous ATLAS 
searches \cite{atlas_7tev_pub_1fb,atlas_7tev_pub,atlas_8tev_pub}. 
The cross-section calculations assume the \wp\ has the same couplings as the SM $W$ boson. The region above each curve is excluded at 95\% CL.
}
  \label{fig:combinedlimits}
\end{figure}

\FloatBarrier

\section{Conclusion}
\label{sec:conclusion}
The ATLAS detector at the LHC has been used to search for new high-mass
states decaying to a lepton plus missing transverse momentum in $pp$ collisions
at $\sqrt{s} = 13$ \tev\ using 3.2 \ifb\ of integrated luminosity.
Events with high-\pt\ electrons and muons and with high \met{} are selected and the transverse mass spectrum is examined. The
data and the SM predictions are in agreement.
Using a Bayesian interpretation, mass limits are set for a possible Sequential Standard Model \wp\ boson. 
Masses below 4.07~\tev\ are excluded at 95\% CL for this model. 
These results represent a significant increase of the mass limit by more than 800~\gev\ compared to the previous ATLAS results based on the Run-1 data.


\section*{Acknowledgements}


We thank CERN for the very successful operation of the LHC, as well as the
support staff from our institutions without whom ATLAS could not be
operated efficiently.

We acknowledge the support of ANPCyT, Argentina; YerPhI, Armenia; ARC, Australia; BMWFW and FWF, Austria; ANAS, Azerbaijan; SSTC, Belarus; CNPq and FAPESP, Brazil; NSERC, NRC and CFI, Canada; CERN; CONICYT, Chile; CAS, MOST and NSFC, China; COLCIENCIAS, Colombia; MSMT CR, MPO CR and VSC CR, Czech Republic; DNRF and DNSRC, Denmark; IN2P3-CNRS, CEA-DSM/IRFU, France; GNSF, Georgia; BMBF, HGF, and MPG, Germany; GSRT, Greece; RGC, Hong Kong SAR, China; ISF, I-CORE and Benoziyo Center, Israel; INFN, Italy; MEXT and JSPS, Japan; CNRST, Morocco; FOM and NWO, Netherlands; RCN, Norway; MNiSW and NCN, Poland; FCT, Portugal; MNE/IFA, Romania; MES of Russia and NRC KI, Russian Federation; JINR; MESTD, Serbia; MSSR, Slovakia; ARRS and MIZ\v{S}, Slovenia; DST/NRF, South Africa; MINECO, Spain; SRC and Wallenberg Foundation, Sweden; SERI, SNSF and Cantons of Bern and Geneva, Switzerland; MOST, Taiwan; TAEK, Turkey; STFC, United Kingdom; DOE and NSF, United States of America. In addition, individual groups and members have received support from BCKDF, the Canada Council, CANARIE, CRC, Compute Canada, FQRNT, and the Ontario Innovation Trust, Canada; EPLANET, ERC, FP7, Horizon 2020 and Marie Sk{\l}odowska-Curie Actions, European Union; Investissements d'Avenir Labex and Idex, ANR, R{\'e}gion Auvergne and Fondation Partager le Savoir, France; DFG and AvH Foundation, Germany; Herakleitos, Thales and Aristeia programmes co-financed by EU-ESF and the Greek NSRF; BSF, GIF and Minerva, Israel; BRF, Norway; Generalitat de Catalunya, Generalitat Valenciana, Spain; the Royal Society and Leverhulme Trust, United Kingdom.

The crucial computing support from all WLCG partners is acknowledged gratefully, in particular from CERN, the ATLAS Tier-1 facilities at TRIUMF (Canada), NDGF (Denmark, Norway, Sweden), CC-IN2P3 (France), KIT/GridKA (Germany), INFN-CNAF (Italy), NL-T1 (Netherlands), PIC (Spain), ASGC (Taiwan), RAL (UK) and BNL (USA), the Tier-2 facilities worldwide and large non-WLCG resource providers. Major contributors of computing resources are listed in Ref.~\cite{ATL-GEN-PUB-2016-002}.





\printbibliography

\newpage 
\begin{flushleft}
{\Large The ATLAS Collaboration}

\bigskip

M.~Aaboud$^{\rm 135d}$,
G.~Aad$^{\rm 86}$,
B.~Abbott$^{\rm 113}$,
J.~Abdallah$^{\rm 64}$,
O.~Abdinov$^{\rm 12}$,
B.~Abeloos$^{\rm 117}$,
R.~Aben$^{\rm 107}$,
O.S.~AbouZeid$^{\rm 137}$,
N.L.~Abraham$^{\rm 149}$,
H.~Abramowicz$^{\rm 153}$
% ...
\bigskip
\\
% ...
$^{12}$ Institute of Physics, Azerbaijan Academy of Sciences, Baku, Azerbaijan\\
% ...
$^{64}$ University of Iowa, Iowa City IA, United States of America\\
% ...
$^{86}$ CPPM, Aix-Marseille Universit{\'e} and CNRS/IN2P3, Marseille, France\\
% ...
$^{107}$ Nikhef National Institute for Subatomic Physics and University of Amsterdam, Amsterdam, Netherlands\\
% ...
$^{113}$ Homer L. Dodge Department of Physics and Astronomy, University of Oklahoma, Norman OK, United States of America\\
% ...
$^{117}$ LAL, Univ. Paris-Sud, CNRS/IN2P3, Universit{\'e} Paris-Saclay, Orsay, France\\
% ...
$^{135}$ $^{(a)}$ Facult{\'e} des Sciences Ain Chock, R{\'e}seau Universitaire de Physique des Hautes Energies - Universit{\'e} Hassan II, Casablanca; $^{(b)}$ Centre National de l'Energie des Sciences Techniques Nucleaires, Rabat; $^{(c)}$ Facult{\'e} des Sciences Semlalia, Universit{\'e} Cadi Ayyad, LPHEA-Marrakech; $^{(d)}$ Facult{\'e} des Sciences, Universit{\'e} Mohamed Premier and LPTPM, Oujda; $^{(e)}$ Facult{\'e} des sciences, Universit{\'e} Mohammed V, Rabat, Morocco\\
% ...
$^{137}$ Santa Cruz Institute for Particle Physics, University of California Santa Cruz, Santa Cruz CA, United States of America\\
% ...
$^{149}$ Department of Physics and Astronomy, University of Sussex, Brighton, United Kingdom\\
% ...
$^{153}$ Raymond and Beverly Sackler School of Physics and Astronomy, Tel Aviv University, Tel Aviv, Israel\\
% ...
$^{a}$ Also at Department of Physics, King's College London, London, United Kingdom\\
$^{b}$ Also at Institute of Physics, Azerbaijan Academy of Sciences, Baku, Azerbaijan\\
$^{c}$ Also at Novosibirsk State University, Novosibirsk, Russia\\
$^{d}$ Also at TRIUMF, Vancouver BC, Canada\\
$^{e}$ Also at Department of Physics {\&} Astronomy, University of Louisville, Louisville, KY, United States of America\\
% ...
\end{flushleft}



\end{document}